\documentclass[doublecol]{epl2}
\title{Searching for top quark pair production cross section at LHeC and FCC-eh }
\author{B.Rezaei\inst{1} \and G.R.Boroun\inst{1}}

\institute{
  \inst{1} Physics Department, Razi University, Kermanshah
67149, Iran} 

\abstract{ The  deep inelastic scattering mode of $t\overline{t}$
pair production at the proposed LHeC and FCC-eh is considered. We
present a method to extract the top reduced cross section related
to the transversal structure function $F_{2}(x,Q^{2})$
parameterization. Numerical calculations with known kinematics of
the LHeC and FCC-eh colliders are demonstrated. The results
obtained for charm and beauty pair production  are comparable with
the experimental data. We show that for a wide range of the
momentum transfer into the top quark pair, the reduced cross
section is well described by center-of-mass energies.}
\begin{document}

\maketitle

\section{1. INTRODUCTION}
The highest center-of-mass energy  in deep inelastic scattering of
electrons on protons at HERA  were reached to $\sqrt{s}\simeq 320~
\mathrm{GeV}$ [1-2]. In recent years, the particle physics
landscape has greatly evolved due to an appearance the project of
Large Hadron electron Collider (LHeC) with the electron-proton
center of mass energy at  $\sqrt{s} \cong 1.3~ \mathrm{TeV}$. It
could be the high-energy $ep/eA$ collider based at CERN [3-4]. The
LHeC energy is about 30 times the center-of- mass energy range of
ep collisions at HERA. The high-luminosity LHC program would be
uniquely complemented by the LHeC, where it was designed in a
extended Conceptual Design Report (CDR) in [3]. The LHeC leads
into the region of high parton densities at low $x$ values where
center-of- mass energy is approximately $1\mathrm{TeV}$. The
kinematic range in the ($x,Q^{2}$) plane of the LHeC for electron
and positron neutral-current (NC) in the perturbative region is
well below $x{\approx}10^{-6}$ and extends up to $Q{\simeq}1~
\mathrm{TeV}$. Also this behavior will be checked at the Future
Circular Collider (FCC) programme which runs to beyond a TeV in
center-of-mass energy [4]. In this collider the FCC-eh with $50~
\mathrm{TeV}$ proton beams colliding with 60 GeV electrons. In
this ep collision the center-of-mass energy reaches $\simeq 3.5~
\mathrm{TeV}$. The LHeC and FCC-eh collisions lead into the region
of high parton densities at small Bjorken $x$. Deep inelastic
scattering measurements at FCC-eh and LHeC will allow the
determination of the parton distribution functions at very small
$x$. These measurements are pertinent in investigations of
lepton-hadron processes at ultra-high energy (UHE)
neutrino astroparticle physics  [4-5].\\
By electron-proton (ep) colliders, the top quark can be produced
in pair in the deep inelastic scattering (DIS) through neutral
current (NC) production. The top quark distribution in
leptoproduction is dominated by the photon-gluon-fusion (BGF)
where the incident a virtual photon interacts with a gluon from
the target nucleon (i.e., Fig.1). The total cross section for
top-pair quark production at the  FCC-eh is $663~\mathrm{fb}$. At
the LHeC, the $t\overline{t}$ production cross sections,
associated with the electron energies $E_{e}=60,~ 140$ and
$300~\mathrm{GeV}$, are $0.023, ~0.120$ and $0.380~\mathrm{pb}$
respectively. These values are smaller than the $\gamma{p}$
collision where the top pair cross section is as large as
$0.700~\mathrm{pb}$ at $E_{e}=60~\mathrm{GeV}$ [6-7]. These
colliders have a broad top physical potential which can be
consulted through Refs.[8-10]. The $t\overline{t}$ production in
DIS  at the LHeC can be used to measure the $tt\gamma$ vertex
where the cross section depends on it. In contrast at the LHC the
vertex is probed through $t\overline{t}\gamma$ production. Indeed,
pair production in DIS is sensitive to the gluon distribution in
proton. The cross sections, in the LHeC and FCC-eh colliders, will
permit a complete unfolding of the heavier quark distributions in
a hugely extended kinematic range of  $Q^{2}$. By using DIS pair
heavier production in an un-accessed range of $Q^{2}$ and $x$, one
can study top component of the structure function. Also the
nonlinear dynamics
must be observed at very low $x$ values ($x<10^{-6}$).\\
The small $x$ range is also relevant to the interactions of cosmic
ultra high energy neutrinos (e.g. the scattering of cosmic
neutrinos from hadrons) which is related to charm and very low $x$
PDFs in comparison with emerging data from the IceCube
Collaboration [11]. The top quark contribution with mass
$172{\pm}0.5~\mathrm{GeV}$ where measured by ATLAS [12] and CMS
[13] is special among all quarks.\\
At low $Q^{2}$ where heavy quarks are not considered as active,
the most standard pQCD scheme for heavy flavors is the fixed
flavor number scheme (FFNS). For $Q^{2}>m^{2}_{h}$ (where $m_{h}$
is the heavy quark mass ), the variable flavor number schemes
(VFNS) have been introduced. For realistic kinematics it has to be
extended to the case of a general- mass- VFNS (GM-VFNS) [14]. In
GM-VFNS on should take into account quark mass, as one of the
ingredients used in this scheme is the replacement of $x$ by the
rescaled variable $\chi$ because
\begin{eqnarray}
\chi=x(1+\frac{4m_{h}^{2}}{Q^{2}}).\nonumber
 \end{eqnarray}
Within the GM-VFNS,  heavy quark densities arise via the
$g{\rightarrow}Q\overline{Q}$ evolution. It would be interesting
to confront with the top distribution at small $x$ in the LHeC and
FCC-eh projects.\\
The layout of the present paper is as follows. After reviewing the
essential features of the heavy quark pair production in section
2, we calculate the production top quark cross section of the
subprocess $\gamma^{*}g{\rightarrow}t\overline{t}$ at the LHeC and
FCC-eh kinematics in this section. To determine our numerical
results, we
 consider heavy quark cross sections predicted by  the proton structure
 function parameterization in section 3. Finally, we give our summary and conclusions in section 4.\\

\section{2. Theory}
In the small-$x$ range, where only the gluon contribution is
dominant, the heavy quark contributions
$F_{k}^{Q\overline{Q}}(x,Q^{2})$ are given by these forms (for
$k=2,L$):
\begin{eqnarray}
F_{2}^{Q\overline{Q}}(x,Q^{2})=C_{2,g}^{Q\overline{Q}}(x,\xi){\otimes}G(x,\mu^{2})\nonumber\\
F_{L}^{Q\overline{Q}}(x,Q^{2})=C_{L,g}^{Q\overline{Q}}(x,\xi){\otimes}G(x,\mu^{2}),
\end{eqnarray}
where $F_{2}^{Q\overline{Q}}$ and $F_{L}^{Q\overline{Q}}$ refer to
the heavy-quarks transversely and longitudinally structure
functions respectively. The $G(x,Q^{2})$ and $g(x,Q^{2})$
represent the gluon momentum distribution and gluon density
respectively, $G(x,Q^{2})=xg(x,Q^{2})$. Here $C_{\{g,k\}}$ are the
coefficient functions at LO and NLO approximation and  $\mu$ is
the mass factorization scale. They are presented in Ref.[15] in
the following form
\begin{eqnarray}
C_{k,g}(z,\zeta)&{\rightarrow}&C^{0}_{k,g}(z,\zeta)+\frac{\alpha_{s}(\mu^{2})}{4\pi}[C_{k,g}^{1}(z,\zeta)\\\nonumber
&&+\overline{C}_{k,g}^{1}(z,\zeta)\ln\frac{\mu^{2}}{m_{c}^{2}}].\nonumber
\end{eqnarray}
The symbol $\otimes$ denotes convolution according to the usual
form, $f(x)\otimes g(x)=\int_{x}^{1}(dy/y)f(y)g(x/y)$ while for
the heavy-quark production the lower limit should be replace by
$ax$ where $a=1+4\xi$ and $\xi=\frac{m_{Q}^{2}}{Q^{2}}$. The deep
inelastic heavy-quarks structure functions related to the reduced
cross section are given by
\begin{eqnarray}
\sigma^{Q\overline{Q}}_r(x,Q^{2})&=&F^{Q\overline{Q}}_2
(x,Q^2)-{\frac{y^2}{Y_+}}F^{Q\overline{Q}}_L (x,Q^2)
\end{eqnarray}
where $y=Q^2/sx$ is the inelasticity with $s$ the ep center of
mass energy squared and $Y_+ =1+(1-y)^2$.  The small $x$
asymptotic behavior of the gluon density can be exploited by the
following form
\begin{eqnarray}
g(x,Q^{2})|_{x\rightarrow0}{\rightarrow}\frac{1}{x^{1+\lambda_{g}}}.\nonumber
\end{eqnarray}
The quantity $1+\lambda_{g}$ is equal to the intercept of the
so-called BFKL Pomeron. Then Eq.(1) can be rewritten as
\begin{eqnarray}
F_{k}^{Q\overline{Q}}(x,Q^{2})&=&G(x,Q^{2})[\int_{x}^{1}\frac{dy}{y}C_{k,g}^{Q\overline{Q}}(y,\xi)y^{+\lambda_{g}}].\nonumber
\end{eqnarray}\\
To summarize and simplify the equations, we describe the following
statement \begin{eqnarray}
f(x){\odot}g(x){\equiv}\int_{x}^{1}(dy/y)f(y)g(y).\nonumber
\end{eqnarray}
Thus, the above equation can be rewritten in the form convenient
for further discussion
\begin{eqnarray}
F_{k}^{Q\overline{Q}}(x,Q^{2})=G(x,\mu^{2})[C_{k,g}^{Q\overline{Q}}(x,\xi){\odot}
x^{\lambda_{g}}].
\end{eqnarray}
The reduced cross section for heavy quarks is expressed in terms
of the gluon distribution as we have it:
\begin{eqnarray}
\sigma^{Q\overline{Q}}_r=G(x,\mu^{2})[C_{2,g}^{Q\overline{Q}}(x,\xi){\odot}x^{\lambda_{g}}
-{\frac{y^2}{Y_+}}C_{L,g}^{Q\overline{Q}}(x,\xi){\odot}x^{\lambda_{g}}].
\end{eqnarray}
At  small $x$ the gluon determination comes from the extension of
range and precision in the measurement of $F_{2}$ and $\partial
F_{2}/\partial \ln Q^{2}$. Several methods of relating the $F_{2}$
scaling violations to the gluon density at small $x$ have been
suggested previously [16]. These relations  estimate the
logarithmic slopes $F_{2}$ with respect to the gluon distribution.
Recently a relation between the gluon distribution and $F_{2}$ and
$\partial F_{2}/\partial \ln Q^{2}$ has been presented in  [17],
with the result
\begin{eqnarray} G(x,Q^{2})&=&
\frac{1}{\Theta_{qg}(x,Q^{2})}[\frac{{\partial}F_{2}(x,Q^{2})}{{\partial}{\ln}Q^{2}}\nonumber\\
&&-\Phi_{qq}(x,Q^{2})F_{2}(x,Q^{2})].
\end{eqnarray}
Indeed the measurement of $F_{2}(x,Q^{2})$ and $\partial
F_{2}(x,Q^{2})/\partial{\ln}Q^{2}$ determine $G(x,Q^{2})$ at low
$x$ kinematic region. The parameterization $F_{2}(x,Q^{2})$
suggested in Ref.[18] by BDH (i.e., M.M.Block, L.Durand and P.Ha)
and also suggested for the longitudinal structure function by KKCZ
(i.e., L.P.Kaptari, A.V.Kotikov, N.Yu.Chernikova and P.Zhang). In
equation (6), kernels for the quark and gluon sectors (denoted by
$\Phi$ and $\Theta$) presented by the following forms
\begin{eqnarray}
\Theta_{qg}(x,Q^{2})&=&P_{qg}(x,\alpha_{s}){\odot}
x^{\lambda_{g}},\nonumber\\
\Phi_{qq}(x,Q^{2})&=&P_{qq}(x,\alpha_{s}){\odot}
x^{\lambda_{s}}\nonumber\\
\end{eqnarray}
where the splitting functions up to NNLO demonstrated  in Ref.[19]
by the following form as:
\begin{eqnarray}
P_{ij}(x,\alpha_{s}(Q^{2}))&=&P_{ij}^{\rm
LO}(x)+\frac{\alpha_{s}(Q^{2})}{2\pi}P_{ij}^{\rm
NLO}(x)\nonumber\\
&&+(\frac{\alpha_{s}(Q^{2})}{2\pi})^{2} P_{ij}^{\rm
NNLO}(x).\nonumber
\end{eqnarray}
 The exponents $\lambda_{s}$ and
$\lambda_{g}$ are defined by the derivatives of the distribution
functions in the form
\begin{eqnarray}
 \lambda_{i}&=&{\partial \ln f^{i}(x,Q^{2})}/{\partial
\ln(1/x)},
\end{eqnarray}
where $i=s,g$ and $f^{s(g)}$ are the singlet structure and gluon
distribution functions respectively. The original behavior for
Eq.(8) was the theoretical expectation at sufficiently small
values of $x$. In Ref.[20] an exponent for the gluon distribution
function at $Q^{2}=1~\mathrm{GeV}^{2}$ for MSTW08 NLO computed
which the fitted value with its uncertainties
 is obtained to be $-0.428^{+0.066}_{-0.057}$ at low values of $x$.
 In addition, the effective exponent values for the gluon distribution at
 $Q^{2}=10
 ~\mathrm{GeV}^{2}$ and $x=10^{-4}$ were evaluated  by means of
 different parameterizations ( NNPDF3.0 [21],  MMHT14 [22], CT14 [23], ABM12 [24] and
  CJ15 [25]). The results are as follows: of  -0.20, -0.15, -0.29, -0.15 and
  -0.14, respectively. For the gluon distribution, the intercept value by the fixed coupling leading log(1/x) BFKL
  solution is defined by
  $\lambda_{g}{\simeq}-0.5$ (which is the so-called hard-pomeron exponent).\\
 At low values of $x$, the transition from a low-$Q^{2}$  to a high-$Q^{2}$
 domain predicted with respect to  the $Q^{2}$ dependence of the effective
 exponent. The asymptotic form these exponents have been predicted
 [20] by the following forms
 \begin{eqnarray}
\lambda_{s}{\rightarrow}-\frac{\gamma}{\rho}+\frac{3}{4\sigma\rho},~~~~
\lambda_{g}{\rightarrow}-\frac{\gamma}{\rho}+\frac{1}{4\sigma\rho}
 \end{eqnarray}
where $\gamma\equiv (\frac{12}{\beta_{0}})^{1/2}$ and
$\beta_{0}=11-\frac{2}{3}n_{f}$ ($n_{f}$ is the active flavor
number). The variables $\sigma$ and $\rho$ are defined as
 \begin{eqnarray} \sigma\equiv [\ln
\frac{x_{0}}{x}\ln\frac{\ln(\frac{Q^{2}}{\Lambda^{2}})}{\ln(\frac{Q_{0}^{2}}{\Lambda^{2}})}]^{1/2}\nonumber
\end{eqnarray}
 and
\begin{eqnarray} \rho\equiv [\frac{\ln
\frac{x_{0}}{x}}{\ln({\ln(\frac{Q^{2}}{\Lambda^{2}})}/{\ln(\frac{Q_{0}^{2}}{\Lambda^{2}})})}]^{1/2}.\nonumber
\end{eqnarray}
The parameters $x_{0}$ and $Q^{2}_{0}$ define the formal
boundaries of the asymptotic region, and $\Lambda$ is the QCD cut-
off parameter for each heavy quark mass threshold as we take the
$n_{f}=4$ for $m_{c}^{2}<\mu^{2}<m_{b}^{2}$ and $n_{f}=5$ for
$m_{b}^{2}<\mu^{2}<m_{t}^{2}$. For singlet structure function an
effective exponent based on HERA combined data and a
phenomenological model, parameterized in Refs.[26] and [27]
respectively. In Ref.[28] these intercepts are determined and
applied to the deep inelastic
lepton nucleon scattering at low values of $x$.\\
Therefore the final improved heavy quark reduced cross section
related to the $F_{2}$ parameterization, is given by
\begin{eqnarray}
\sigma^{Q\overline{Q}}(x,Q^{2})&=&[C_{2,g}^{Q\overline{Q}}(x,\xi){\odot}x^{\lambda_{g}}
-\frac{y^2}{Y_+}C_{L,g}^{Q\overline{Q}}(x,\xi){\odot}x^{\lambda_{g}}]\nonumber\\
&&\times[\frac{1}{\Theta_{qg}(x,\mu^{2})}\frac{{\partial}F_{2}(x,\mu^{2})}{{\partial}{\ln}\mu^{2}}\nonumber\\
&&-\frac{\Phi_{qq}(x,\mu^{2})}{\Theta_{qg}(x,\mu^{2})}F_{2}(x,\mu^{2})].
\end{eqnarray}
The explicit expression for the $F_{2}$ parameterization, which
suggested by BDH in Ref.[18] obtained in a wide range of the
kinematical variables $x$ and $Q^{2}$ from a combined fit of the
H1 and ZEUS data read as
\begin{eqnarray}
F^{\gamma p}_{ 2}(x,Q^{2})& =& D(Q^{2})(1-
x)^{n}\sum_{m=0}^{2}A_{m}(Q^{2})L^{m},
\end{eqnarray}
where the parameters  with their statistical errors are given in
Ref.[18]. Eq.(10) with respect to range of Eq.(11) cover the
effective behavior of the heavy-quark structure functions in DIS.
Coefficients of the fitted $F_{2}$ are obtained as functions of
$x$ and $Q^{2}$ by using of the HERA data. Consequently, we obtain
the top reduced cross section by using the $F_{2}$
parameterization which extended smoothly and reasonably to values
of $Q^{2}\geq 10^{4}~\mathrm{GeV}^{2}$. These data are needed in
investigations of
ultra-high energy processes.\\
\section{3. Results}
The production of top quarks in ep collisions  at LHeC and FCC-eh
can be  provided a stringent test of new physics at UHE. Test of
pQCD in ep collisions at HERA were provided by production of charm
and beauty quarks in neutral current deep inelastic
electron-proton scattering. Previous measurements [29] at HERA
have demonstrated that charm and beauty quarks are produced via
the boson-gluon fusion process which they are sensitive to the
gluon density in the proton and the heavy-quarks mass. The
heavy-quarks masses are set to $m_{c}=1.5{\pm}0.15~\mathrm{GeV}$
and $m_{b}=4.5{\pm}0.25~\mathrm{GeV}$. The charm and beauty
structure functions are obtained from the measured cross sections
in [29] and studied in [30-31] phenomenological successfully in
recent years. The b-quark density is important in Higgs production
at the LHC. Also the t-quark density will be important to study
the Higgs boson at the LHeC and FCC-eh in UHE and nonlinear gg
interaction
effects [2] at very low values of $x$.\\
In Fig.2, phenomenological predictions of the charm and beauty
reduced cross sections are compared to the combined HERA data
[29]. The renormalisation and factorisation scale for the heavy
quarks is set to $<\mu^{2}>=4m_{Q}^{2}+\frac{Q^{2}}{2}$. The
center-of-mass energy ($\sqrt{s}=$318~GeV) for charm and
beauty-quark production used in the combined HERA data. In this
figure the charm and beauty reduced cross sections are determined
with respect to the $F_{2}$ parameterization [18] and compared to
the results of the HERA combined [29] at
$Q^{2}=120~\mathrm{GeV}^{2}$. These reduced cross sections
determined using a strong coupling constant
$\alpha_{s}^{n_{f}=3}(M_{Z})=0.105{\pm}0.002$ which correspond to
$\alpha_{s}^{n_{f}=5}(M_{Z})=0.116{\pm}0.002$. The uncertainty of
the reduced charm and beauty cross sections are due to the $F_{2}$
parameterization, singlet exponent  and mass uncertainties.
Consistency between the determined results with respect to the
experimental data can be observed. These results are also
comparable with results in Refs.[30-31]. We observe that the
uncertainties of $\sigma_{r}^{b\overline{b}}$ are lower than the
experimental uncertainties. Here there are two reasons for this
process. Usually, the data obtained for $b\overline{b}$
pair-production in DIS have larger uncertainties than
$c\overline{c}$ pair-production. Also, the $F_{2}$
parameterization is based on $n_{f}=4$ as suggested by BDH in
Ref.[18].\\
Now we focus attention on the phenomenological prediction for the
top quark production  in the LHeC and FCC-eh collisions with
center-of-mass energy $1.3~\mathrm{TeV}$ and $3.5~\mathrm{TeV}$
respectively. The $t\overline{t}$ production at the Tevatron
collider and  LHC discussed   in Refs.[14] and [28] at NNLO. Total
cross section for top quark production at $t\overline{t}$
photoproduction is $1.14~\mathrm{pb}$ as reported in Refs.[2-4].
Determination of $\alpha_{s}$ at the LHeC is according to the H1
result at NNLO
($\alpha_{s}(M_{z}^{2})=0.1157{\pm}0.0020(exp.){\pm}0.0029(thy.)$)
with $0.2\%$ uncertainty  from the LHeC and $0.1\%$ when combined
with HERA [2]. Here we note that the LHeC uncertainties are
simulated [2-4]. The singlet and  gluon exponents are determined
in accordance with data in Refs.[20] and [32].  The average value
of the parameter $y$ was chosen equal to $<y>=0.5$, since the
minimum and maximum values of the  top reduced cross section are
determined by the inelasticity $y=1$ and $y=0$ respectively. Fig.3
shows the theoretical prediction for $\sigma^{t\overline{t}}_{r}$
as a function of $x$ using the parameterization of
$F_{2}(x,Q^{2})$. The solid curves are correspondent to the scale
choice $Q^{2}=\frac{1}{4}m_{t}^{2}$, $Q^{2}=m_{t}^{2}$ and
$Q^{2}=4m_{t}^{2}$. Model uncertainties arise from the variations
of the $F_{2}$ parameterization, top quark mass and the singlet
exponent behavior. Our numerical results as accompanied with the
statistical errors are summarized in this figure(i.e., Fig.3).
Here  the top reduced cross sections are plotted as a function of
$x$ at values of
$\frac{1}{4}m_{t}^{2}{\leq}Q^{2}{\leq}m_{t}^{2}~[\mathrm{GeV}]^{2}$.
We observe that the reduced cross sections
,$\sigma_{r}^{t\overline{t}}$, are in the range of $0.01\sim0.4$
as $x$ decreases.\\
In Fig.4, we compared the results of top reduced cross section for
center-of-mass energies $\sqrt{s}=1.3$ and $3.5~\mathrm{TeV}$
separately. In this figure $\sigma_{r}$ plotted for $100 \leq
Q^{2}<4m_{t}^{2}~[\mathrm{GeV}^{2}]$ and it is  assumed that
inelasticity is constant in this process, $y=0.5$. At fixed center
of mass energy, $\sqrt{s}$, the variables are related by the
following rewritten form based on the rescaled variable $\chi$ as
$Q^{2}=s{\chi}y$. The effects of $y$ constant for the reduced
cross sections have bee shown in this figure. In $Q^{2}$ range  a
enhancement is observable until $Q^{2}{\simeq}m^{2}_{t}$. Then a
depletion is observable, because for $Q^{2}{>}m^{2}_{t}$ we do not
expect that the inelasticity to be $0.5$. The validity of these
results to the top reduced cross section
could be checked in the future at the proposed LHeC and FCC-eh colliders.\\

\section{4. Summary and Conclusion}
We have studied the production of top-pair quarks in new electron
proton collisions (i.e., LHeC and FCC-eh). The subprocess
$\gamma^{*}g{\rightarrow}t\overline{t}$ will be one kind of
important production channels at LHeC and FCC-eh. The production
of charm and beauty quarks were studied in the basic processes of
$c\overline{c}$ and $b\overline{b}$ production at HERA. The method
rely on the DGLAP evolution equations and the proton structure
function parameterization. We focus on the kinematic region of
low-$x$ and high-$Q^{2}$ values which proposed at new colliders.
The obtained explicit expression for $\sigma_{r}^{Q\overline{Q}}$
is entirely determined by the $F_{2}^{BDH}$ parameterization which
extended to values of high-$Q^{2}$. The results of numerical
calculations for charm and beauty as well as comparisons with
available experimental data are presented. We considered the top
reduced cross section behavior at low values of $x$ in a wide
range of $Q^{2}$
values.\\


\begin{figure}
\includegraphics[width=0.40\textwidth]{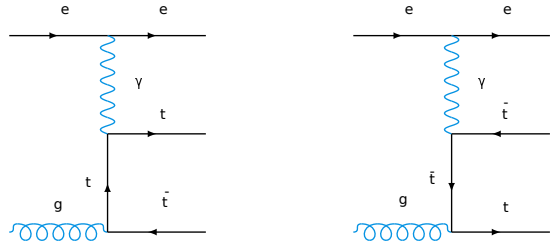}
\caption{Deep inelastic ep scattering due to boson gluon
fusion.}\label{Fig1}
\end{figure}
\begin{figure}
\includegraphics[width=0.5\textwidth]{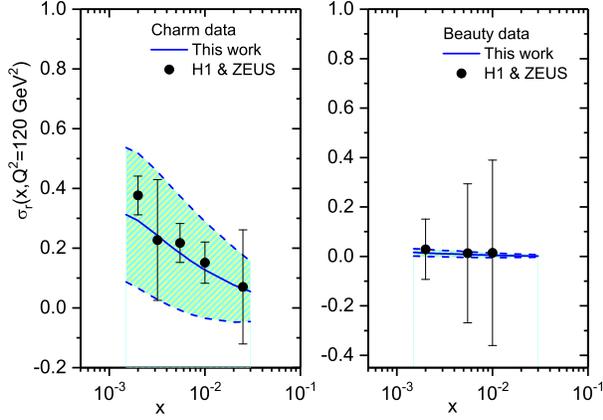}
\caption{The charm and beauty components of the reduced cross
section given by $\sigma_{r}^{c\overline{c}}$ and
$\sigma_{r}^{b\overline{b}}$ as a function of $x$ at
$Q^{2}=120~\mathrm{GeV}^{2}$ as accompanied with statistical
errors. Experimental data are from the H1 and ZEUS Collaborations,
Ref.[29]. }\label{Fig1}
\end{figure}
\begin{figure}
\includegraphics[width=0.5\textwidth]{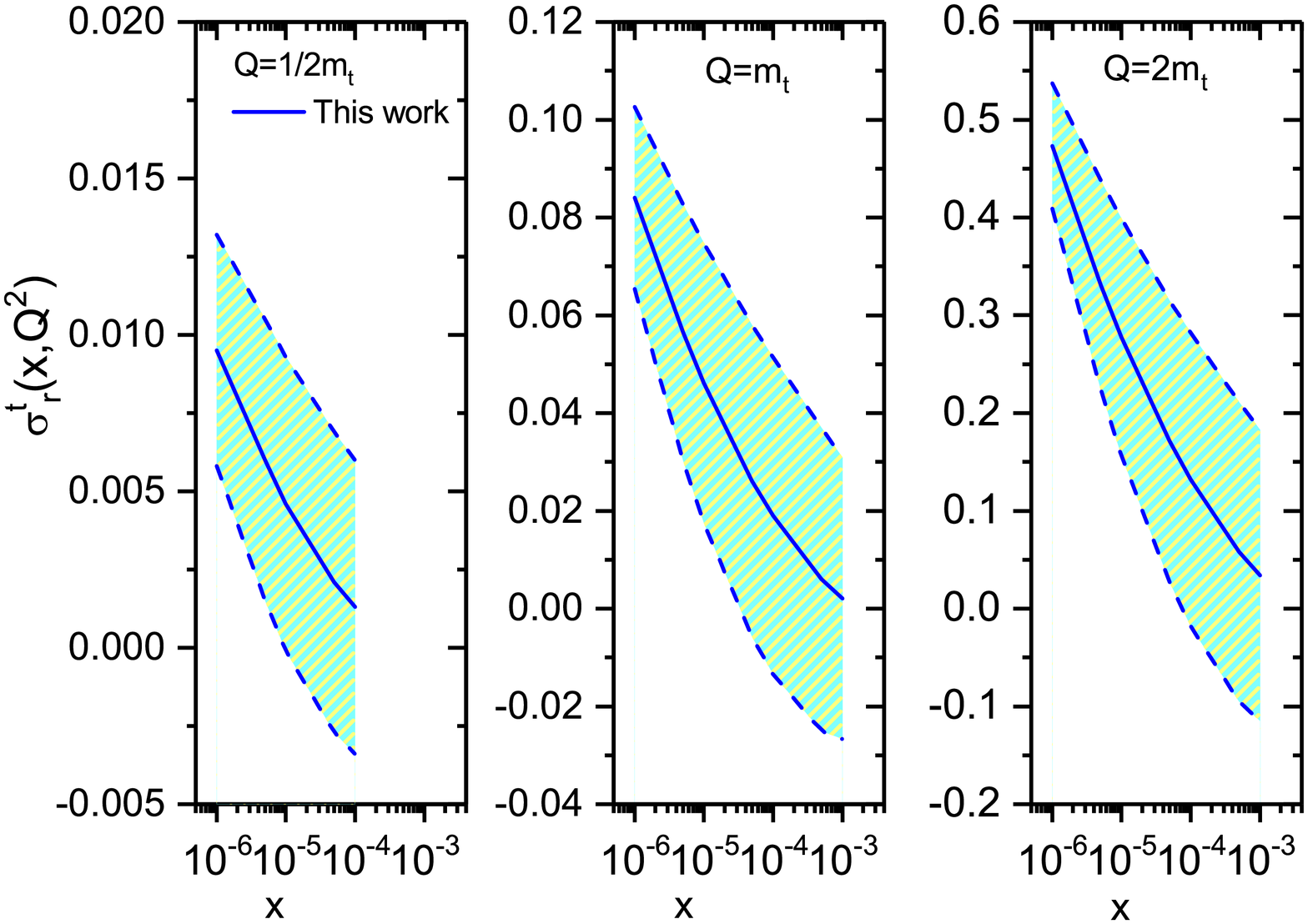}
\caption{Theoretical predictions for $\sigma^{t\overline{t}}_{r}$
at $y=0.5$ as a function of $x$ using the $F_{2}$
parameterization. The curves are calculated using
$Q^{2}=\frac{1}{4}m_{t}^{2}$, $m_{t}^{2}$ and
$4m_{t}^{2}~[\mathrm{GeV}^{2}]$ as accompanied with statistical
errors.}\label{Fig1}
\end{figure}
\begin{figure}
\includegraphics[width=0.5\textwidth]{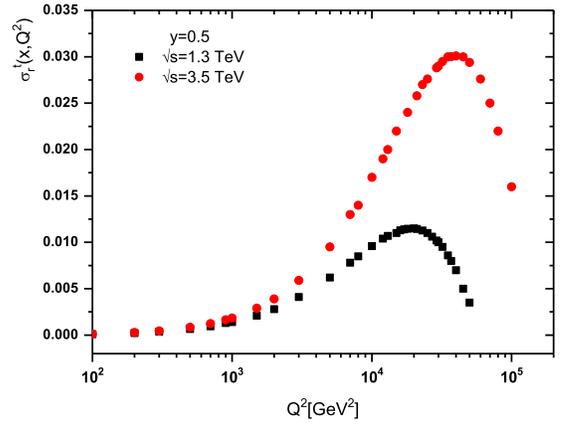}
\caption{Comparison of the results obtained for the top reduced
cross section, $\sigma^{t\overline{t}}_{r}$, at center-of-mass
energies $\sqrt{s}=1.3$ and $3.5~\mathrm{TeV}$ as a function of
$Q^{2}$ values. In these processes, $y$ is equal to
$0.5$.}\label{Fig1}
\end{figure}


%

\end{document}